\documentclass[english, 11pt]{article}
\usepackage[T1]{fontenc}
\usepackage[latin9]{inputenc}
\usepackage{babel, hyperref, graphicx, cite}
\begin{document}

\title{Evaluation of crystal free energy with lattice dynamics}

\author{Tran Doan Huan\\{\it Institute of Materials Science, University of Connecticut, Storrs, CT 06269-3136, USA}\\
{\it Email: huan.tran@uconn.edu}}
\maketitle

\begin{abstract}
Within the framework of density functional theory (DFT), the total energy of crystal structures is calculated at zero temperature. Herein, we briefly discuss the DFT-based lattice-dynamics approach for computing crystal free energy, the quantity needed in various non-zero-temperature contexts. We illustrate this well-established approach by examining the temperature-dependent thermodynamic stability of several crystalline materials, including ZrO$_2$, HfO$_2$, KBH$_4$, and Zn(BH$_4$)$_2$.
\end{abstract}

\section{Introduction}
Density functional theory (DFT) \cite{DFT1, DFT2,MartinDFT} is a powerful numerical method in various disciplinaries, e.g., computational chemistry and computational materials science. In a typical calculation with DFT, the total energy of a {\it static} atomic (or more precisely, ionic) structure $\{{\bf r}_i\}$ --- here ${\bf r}_i$ is the coordinate of the $i^{\rm th}$ ion --- of a solid structure of $N$ atoms, is evaluated. The forces exerting on these ions can also be computed with the Hellmann-Feynman theorem\cite{FeynmanForce} and then be used to relax the structure, i.e., to locate a local minimum of the potential energy surface in the configurational space. Some materials properties, e.g., band gap, can then be calculated on top of the relaxed structures obtained. Because temperature is not included in the DFT formalism, all of these properties, when calculated with DFT, are valid at zero temperature ($T=0$). Of particular interest are, however, certain materials properties at some sort of finite temperatures, i.e., $T>0$. Technically, it is possible to evaluate some properties at non-zero temperatures, of course, with certain computational overhead.

This contribution describes a method, which is based on lattice dynamics \cite{BornHuang}, for estimating the free energy of crystals at non-zero temperatures. Presumably, this is one of the most desirable finite-temperature quantities as it determines pretty much the essential characters of a crystal \cite{BornHuang,Pathria,KuhnBook}. To be more precise, the free energy difference between systems and the gradients of the free energy with respect to thermodynamic variables are of practical interest. For instance, among various possible crystal structures, that with lowest-possible free energy is considered to be thermodynamically stable. For a chemical reaction, the free energy change is an useful parameter, indicating whether the reaction is spontaneous or not, and if not, how much energy is needed to make it occur.

The Gibbs free energy of an equilibrium (relaxed) structure $\{{\bf r}^0_i\}$ at constant number of particles is defined as $G(P,V,T) \equiv U - TS + PV$ where $U$ is the internal energy, $S$ is the entropy, $P$ is the pressure, and $V$ is the volume (per unit cell, for examples) of the crystals. Strictly, the ions are not frozen but perform certain vibrations in the neighborhood of $\{{\bf r}^0_i\}$. For this reason, the internal energy may be expressed as $U = U_0 + U_{\rm vib}$ where $U_0$ is the static internal energy at $\{{\bf r}^0_i\}$ and $U_{\rm vib}$ is the vibrational internal energy \cite{KuhnBook}. Technically, $U_0$ is the total DFT energy of the relaxed structure $\{{\bf r}^0_i\}$ while $PV$ can also readily be computed. The treatment for the remaining terms, i.e., the vibrational free energy defined as $F_{\rm vib}(T)\equiv U_{\rm vib}-TS$, which is considered in details herein, is based on the calculations of the vibrational frequency spectra of the examined solids. This approach is well-established and has widely been used \cite{Herbst,VossMgBH4,Huan:Zn,Huan:Alanates,Huan:NaSc,huan:hafnia,Tuan:PSSB,PtSO4,C4CP05644B}, showing some advantages and shortcomings which will also be discussed in this contribution.

\section{Vibrational free energy}
To focus on $F_{\rm vib}(T)$, we put the $PV$ term of $G$ aside, and consider the Helmholtz free energy $F(T) \equiv U - TS = U_0 + F_{\rm vib}(T)$. The formal definition of $F(T)$ is\cite{BornHuang,Pathria}
\begin{equation}\label{eq:F}
F(T) \equiv -k_{\rm B}T \ln Z
\end{equation}
where $k_{\rm B}$ is the Boltzmann's constant, and $Z$ is the canonical partition function defined by
\begin{equation}\label{eq:Z}
Z = \int d\{{\bf r}_i\} \exp\left[-\frac{H(\{{\bf r}_i\})}{k_BT}\right].
\end{equation}
Here, the integral is taken over all the possible atomic configuration $\{{\bf r}_i\}$ of the system, and $H = T + \Phi$ where $T$ is the kinetic energy of the ionic lattice and $\Phi\equiv \Phi(\{{\bf r}_i\})$ is the potential of the ion structure. In principles, the very-high-dimensional integral appearing in Eq. (\ref{eq:Z}) can not be performed directly. It may, however, be sampled from a thermodynamic ensemble by a number of techniques, many of them are based on Monte Carlo like the Bennett acceptance ratio method \cite{BAR}. For a better summary of such methods, readers are referred to Ref. \cite{Ackland:freeenergy}.

The lattice-dynamics approach for $F_{\rm vib}(T)$ is started from the harmonic approximation of the potential energy $\Phi(\{{\bf r}_i\})$, i.e., \cite{BornHuang}
\begin{equation}\label{Eq:H1}
\Phi(\{{\bf r}_i\}) = \Phi_0(\{{\bf r}^0_i\})+\frac{1}{2}\sum_{i,j,\alpha,\beta}\frac{\partial^2 \Phi(\{{\bf r}_i\})}{\partial r_{i,\alpha}\partial r_{j,\beta}}u_{i,\alpha}u_{j,\beta}.
\end{equation}
In this expansion, $u_{i,\alpha}$ ($u_{j,\beta}$) is the displacement of ion $i$ ($j$) along the $\alpha$ ($\beta$) direction. It should be noted here that $\Phi_0=U_0$, the internal energy of the frozen ionic structure $\{{\bf r}^0_i\}$ and $U_{\rm vib} = T + \frac{1}{2}\sum_{i,j,\alpha,\beta}\frac{\partial^2 \Phi(\{{\bf r}_i\})}{\partial r_{i,\alpha}\partial r_{j,\beta}}u_{i,\alpha}u_{j,\beta}$. By (\ref{Eq:H1}), the displacements are assumed to be small, and hence unharmonic terms like those arisen from substructure rorations and the translations (these motions may be possible in some highly dynamical solids like complex borohydrides \cite{DF9551900230,BH4_rotate,LiBH4_rotate}) are not included. In the quantum description of the lattice vibrations, a set of normal coordinates $\{q_\nu\}$ and $p_\nu=-i\hbar\partial/\partial q_\nu$ ($\hbar$ is the Planck constant) are introduced so that \cite{BornHuang}
\begin{equation}
H = U_0 + T + \frac{1}{2}\sum_{i,j,\alpha,\beta}\frac{\partial^2 \Phi}{\partial r_{i,\alpha}\partial r_{j,\beta}}u_{i,\alpha}u_{j,\beta} = U_0 + \frac{1}{2}\sum_{\nu=1}^{3N}p_\nu^2 + \frac{1}{2}\sum_{\nu=1}^{3N}\omega_\nu^2q_\nu^2
\end{equation}
where $\{\omega_\nu\}$ are the vibrational frequencies. The eigenenergies of $H$ are then
\begin{equation}
\epsilon_n = U_0 + \sum_{\nu=1}^{3N}\left(n_\nu+\frac{1}{2}\right) \hbar\omega_\nu
\end{equation}
in which $n=\{n_1, n_2, ..., n_{3N}\}$. By inserting $\epsilon_{n}$ into (\ref{eq:Z}), one gets $Z=\exp(-U_0/k_{\rm B}T)Z_{\rm vib}$ with
\begin{equation}
Z_{\rm vib}=\prod_{\nu=1}^{3N}\sum_{n_\nu=0}^\infty\exp\left[-\frac{(n_\nu+1/2)\hbar\omega_\nu}{k_{\rm B}T}\right]=\prod_{\nu=1}^{3N}\left[2\sinh\left(\frac{\hbar\omega_\nu}{2k_{\rm B}T}\right)\right]^{-1}.
\end{equation}
Eq. (\ref{eq:F}) implies that $F_{\rm vib}(T)$ is determined by $F_{\rm vib}(T) = -k_{\rm B}T \ln Z_{\rm vib}$, or
\begin{equation}
F_{\rm vib}(T) = k_{\rm B}T\sum_{\nu=1}^{3N}\ln\left[2\sinh\left(\frac{\hbar\omega_\nu}{2k_{\rm B}T}\right)\right].
\end{equation}
At the limit of $T\to 0$, $F_{\rm vib}(T)$ approaches the zero-point energy, which is given by
\begin{equation}
E_{\rm ZP} = \sum_{\nu=1}^{3N}\frac{\hbar\omega_\nu}{2}.
\end{equation}
In case the spectrum of $\omega$ is continuous, $F_{\rm vib}(T)$ is written in terms of the normalized density of phonon states $g(\omega)$ as
\begin{equation}\label{eq:fvib}
F_{\rm vib}(T) = 3Nk_{\rm B}T\int_0^\infty d\omega g(\omega)\ln\left[2\sinh\left(\frac{\hbar\omega}{2k_{\rm B}T}\right)\right].
\end{equation}
Thus, given that the density of phonon states $g(\omega)$ is determined by whatever method, the vibrational free energy $F_{\rm vib}(T)$ can readily be calculated.

\section{Calculations of $g(\omega)$ with DFT}
The density of phonon states $g(\omega)$ can be calculated numerically by some methods. Here, we examine the prescriptions made available within the DFT-based framework, leaving aside other methods, e.g., that starts from the velocity autocorrelation function which can be accessible with molecular dynamics.

Calculations of $g(\omega)$ with DFT involve the determination of phonon dispersion relation $\omega_\nu({\bf q})$ --- here $\bf q$ is the phonon wave vector --- by diagonalizing the dynamical matrix, defined as\cite{BornHuang}
\begin{equation}\label{eq:D}
D_{\alpha\beta}(ij; {\bf q}) = \frac{1}{\sqrt{m_im_j}} \sum_l \frac{\partial^2 \Phi}{\partial r_{i,\alpha}\partial r_{j,\beta}}\exp\left(i{\bf q}\cdot{\bf r}_l\right)
\end{equation}
where $m_i$ and $m_j$ are the mass of ions $i$ and $j$, respectively. The sum in (\ref{eq:D}) is taken over the index $l$ of the translational images of the crystal cell due to the periodicity. With the dynamical matrix $D$ introduced, the equation of motion of the vibrations is written as
\begin{equation}
\sum_{j\beta}D_{\alpha\beta}(ij;{\bf q})e_\beta(j, {\bf q}\nu)=m_i\omega^2_\nu({\bf q})e_\alpha(i,{\bf q}\nu)
\end{equation}
where $e_\alpha(i,{\bf q}\nu)$ are the coefficients of the transformation from $u_{i,\alpha}$ to $q_\nu$ \cite{BornHuang}. Thus, the problem of calculating the phonon dispersion $\omega_\nu({\bf q})$ of the $\nu^{\rm th}$ normal phonon mode is reduced to the eigenvalue problem of $D_{\alpha\beta}(ij;{\bf q})$, which can be solved numerically.

In practice, $D_{\alpha\beta}(ij; {\bf q})$ and $\omega_\nu({\bf q})$ are computed by either the frozen-phonon method (also referred to as the direct approach) \cite{phonopy_sc} or the linear-response method \cite{DFPT,Abinit_phonon_1,Abinit_phonon_2}. The former relies on evaluating the interatomic force constants of a large supercell in which the ions are perturbed from their equilibrium positions. The size of the supercell may be large, depending on the commensurability of the perturbations with the equilibrium cell. Then, the obtained matrix of force constants is reduced to the dynamical matrix at certain values of $\bf q$ which is then be diagonalized to obtain $\omega_\nu({\bf q})$. In this method, a DFT package like {\sc vasp}\cite{vasp1,vasp2,vasp3,vasp4} or {\sc siesta} \cite{SIESTA} is needed as a force calculator, while an additional software, e.g., {\sc phonon} \cite{phonopy_sc}, {\sc phonopy} \cite{phonopy}, and {\sc phon} \cite{PHON}, is used for the pre- and post-processing calculations.

The linear-response method for calculating $\omega_\nu({\bf q})$ and $g(\omega)$ is based on the perturbation theory within DFT\cite{DFPT}. In this method, the dynamical matrix is evaluated from the knowledge of the first-order response of the wave function with respect to the lattice potential perturbations \cite{Abinit_phonon_1,Abinit_phonon_2}. Different from the frozen-phonon method which needs supercell to be constructed, the dynamical matrix is {\it separately} evaluated by linear-response calculations on the unit cell at various $\bf q$ points. Available packages that support the linear-response method for calculations of $\omega_\nu({\bf q})$ are, but not limited to, {\sc abinit} \cite{Gonze_Abinit_1} and {\sc quantum espresso} \cite{QE}. 

Given that the phonon dispersion $\omega_\nu({\bf q})$ is determined, the density of phonon states $g(\omega)$ and the vibrational free energy $F_{\rm vib}(T)$ can be straightforwardly calculated, as supported by all of the codes/tools mentioned above. In addition, calculated $\omega_\nu({\bf q})$ is also very useful to examine the dynamical stability of crystal structures, especially those predicted theoretically \cite{Huan:Zn,Huan:Alanates,Huan:NaSc,huan:hafnia,PtSO4,C3RA43617A}. In particular, if a structure corresponds to a saddle point of the potential energy surface, there are at least one phonon modes with imaginary frequencies, and one can follow these modes to end up at one or more lower-energy, dynamically stable structures \cite{Huan:NaSc}.

\section{Free energy and thermodynamic stability}\label{sec:therm}
In this Section, we examine two examples of using the calculated free energy for some discussions of the thermodynamic stability of crystals.

\subsection{Temperature-driven structural phase transition}
At a given temperature/pressure condition, the lowest Gibbs free energy structure of a certain crystal is considered to be thermodynamically stable. Consequently, the structural phase transitions of a crystal may be traced by examining the Gibbs free energy calculated for its possible structures \cite{huan:hafnia,Tuan:PSSB,PtSO4,C4CP05644B}. Here, we limit at $P=0$, thus the Helmholtz free energy $F(T)$ will be used to estimate the critical temperature $T_{\rm C}$ of the phase transitions of some representative crystals.

\begin{table}[b]
\begin{center}
\caption{Low- and high-temperature phases of some crystalline materials examined in this work.}\label{table:struct}
\begin{tabular}{lcccc}
\hline
\hline
Materials & Low-T structure & High-T structure & $T_{\rm C}$ & Ref.\\
\hline
ZrO$_2$ & $P2_1/c$ & $P4_2/nmc$ & 1500K & \cite{Ondik}\\
HfO$_2$ & $P2_1/c$ & $P4_2/nmc$ & 2200K & \cite{Ondik}\\
KBH$_4$ & $P4_2/nmc$ & $Fm\overline{3}m$ & 197K &\cite{Renaudin200498}\\
\hline
\end{tabular}
\end{center}
\end{table}

\begin{figure}[t]
\begin{center}
\includegraphics[width=8.5cm]{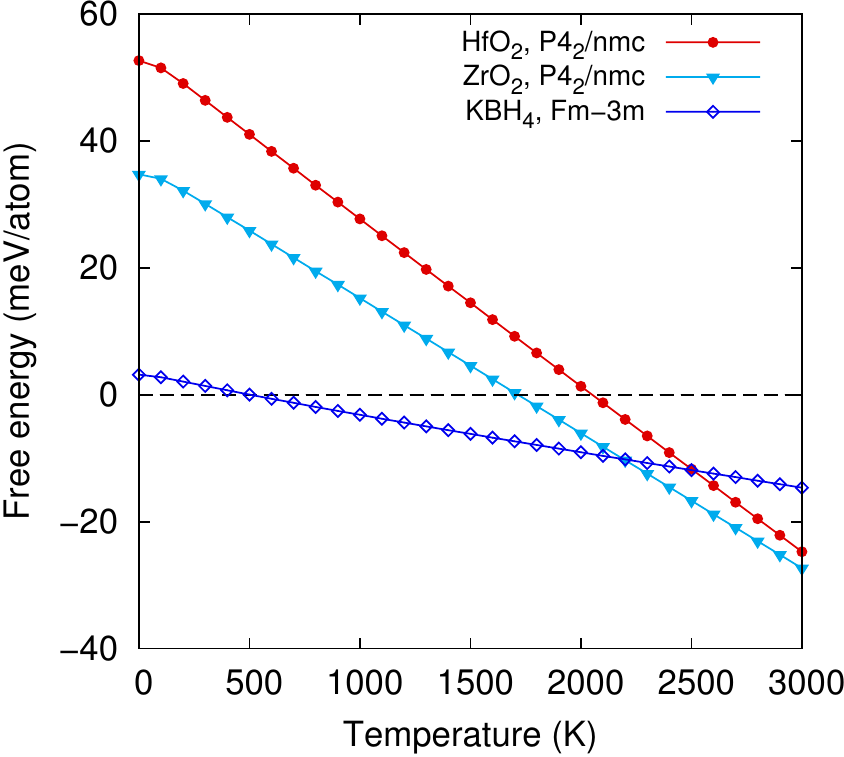}
\caption{Helmholtz free energies calculated for the high-$T$ phases of ZrO$_2$, HfO$_2$ and KBH$_4$ with respect to their low-$T$ phases. Details on these phases can be found in Table \ref{table:struct}.}\label{fig:fe}
\end{center}
\end{figure}

Three crystalline solids which are used for this demonstration, are summarized in Table \ref{table:struct}. Both zirconia ZrO$_2$ and hafnia HfO$_2$ are known\cite{Ondik,Ohtaka:hx5029,Ohtaka:hafnia,Wang_review} to transform from a monoclinic $P2_1/c$ structure at low temperatures to a tetragonal $P4_2/nmc$ structure at higher temperatures. These phase transitions occur at $T_{\rm C}\simeq 1500$K and $T_{\rm C}\simeq 2200$K for zirconia and hafnia, respectively. Likewise, potassium borohydride KBH$_4$ crystallizes in a tetragonal $P4_2/nmc$ structure below $T_{\rm C}\simeq 197$ K while adopting a cubic $Fm\overline{3}m$ structure above this point \cite{Renaudin200498}. The Helmholtz free energy $F(T)$ has been calculated for all of these structural phases, using {\sc vasp} and {\sc phonopy}. In Fig. \ref{fig:fe} we show the free energy of the high-temperature (high-$T$) structure of each crystal with respect to that of the low-temperature (low-$T$) structure. Clearly, the low-$T$ and high-$T$ structures of zirconia, hafnia, and potassium borohydride are correctly described by the calculations. The phase transitions of zirconia and hafnia are predicted to be $1600$K and $2100$ K \cite{huan:hafnia}, which are in relatively good agreements with the known facts. On the other hand, the critical temperature predicted for potassium borohydride is about $550$ K, which is considerably higher than the correct value (a similar observation  was also reported in Ref. \cite{Tuan:PSSB}).

The discrepancy observed for $T_{\rm C}$ of potassium borohydride may be understandable, given that the lattice-dynamics approach for calculating $F_{\rm vib}(T)$ assumes a series of approximations and simplifications. Moreover, as mentioned above, complex borohydrides like KBH$_4$ are highly dynamical, i.e., the BH$_4$ groups may jump and/or rotate \cite{DF9551900230,BH4_rotate,LiBH4_rotate}. Such the motions can not be captured by the harmonic approximation, which assumes small displacements. Next, the lattice-dynamics approach, as described in this contribution, is basically an extrapolation procedure, i.e., $U_0$ and $g(\omega)$ of the high-$T$ cubic $Fm\overline{3}m$ phase, which exists at $T\simeq 197$K, are calculated at $0$K and then, the relevant energetic quantity, i.e., $F(T)$, is extrapolated back to non-zero temperatures using Eq. (\ref{eq:fvib}). Finally, the energy difference between the high-$T$ and low-$T$ phases of potassium borohydride is very small ($\simeq 3$ meV/atom, more than one order smaller than those of zirconia and hafnia), suggesting that this numerical scheme may still not be adequate for a marginal case like KBH$_4$.

\subsection{Thermodynamic stability w.r.t. formation/decomposition}
For another demonstration, we consider a chemical reaction of which the reactants and products are all solids \cite{JeonZnBH4,SrinivasanZnBH4}
\begin{equation}\label{eq:znbh4}
2{\rm NaBH}_4+{\rm ZnCl}_2 \to {\rm Zn(BH_4)}_2+2{\rm NaCl}.
\end{equation}
This is the synthesis route of ${\rm Zn(BH_4)}_2$, as reported in Refs. \cite{JeonZnBH4,SrinivasanZnBH4}. However, the stability of ${\rm Zn(BH_4)}_2$ is still in a debate, and a the presence of pure ${\rm Zn(BH_4)}_2$ is not conclusive. While the structure of the synthesized ${\rm Zn(BH_4)}_2$ samples has yet been resolved, a number of low-energy structures have been proposed via DFT calculations \cite{Nakamori:Zn,Choudhury:Zn,Huan:Zn}. As one of the simple tests that may be performed on these hypothetical structures, the free energy change $\Delta F(T)$ of the reaction (\ref{eq:znbh4}) can be calculated. A positive value of $\Delta F(T)$ suggests that the examined structure of ${\rm Zn(BH_4)}_2$ is unstable, while a negative value of $\Delta F(T)$ may be a supporting evidence for ${\rm Zn(BH_4)}_2$.

\begin{figure}[t]
\begin{center}
\includegraphics[width=8.5cm]{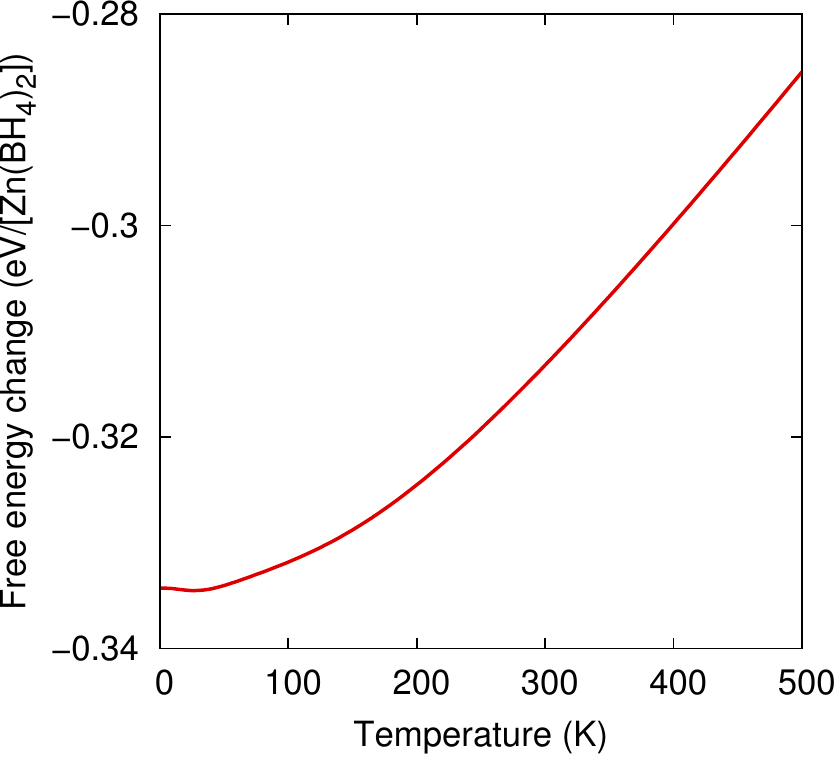}
\caption{Free energy change $\Delta F(T)$ of reaction (\ref{eq:znbh4}).}\label{fig:zn}
\end{center}
\end{figure}

We show in Fig. (\ref{fig:zn}) the free energy change $\Delta F(T)$ calculated for the reaction (\ref{eq:znbh4}), using the currently known most-stable structures of the crystals involving. They are the $I4_122$ structure of ${\rm Zn(BH_4)}_2$ \cite{Huan:Zn}, the $P4_2/nmc$ structure of NaBH$_4$ \cite{Renaudin200498}, the $Fm\overline{3}m$ structure of NaCl, and the $Pna2_1$ structure of $\delta-$ZnCl$_2$ \cite{ZnCl2}. A conclusion which may be drawn from Fig. (\ref{fig:zn}) is that up to $500$ K, the formation of ${\rm Zn(BH_4)}_2$ according to the reaction (\ref{eq:znbh4}) is energetically permissible. It is however worth noting that the thermodynamic stability of ${\rm Zn(BH_4)}_2$ should be examined with respect to all the possible formation/decomposition chemical reactions, the practice which is well beyond the scope of this contribution.

\section{Conclusions}
In summary, Gibbs and Helmholtz free energies of crystals may be estimated by lattice dynamics within the harmonic approximation at the level of density functional theory. This computational scheme produces reasonable results for many crystals but there are also other solids with which this approach should be used with great cautions. Although some treatments for the anharmonic terms of the free energies are also currently available, they should also be material-dependent in the similar fashion with this scheme.


\end{document}